\documentstyle[preprint,epsfig,aps]{revtex}

\def\be{\begin{equation}}
\def\ee{\end{equation}}
\def\bea{\begin{eqnarray}}
\def\eea{\end{eqnarray}}

\begin{document}

%\twocolumn
\tightenlines
\draft

\preprint{\vbox{\hbox{UOM/NPh/HQP/00-3}\hbox{UH-511-962-00}}}

\title{Inclusive charmless semileptonic decay of $\Lambda_b$
and $Br(b \rightarrow X_u l \nu_l)$}

\author{Somasundaram Arunagiri$^a$ and Hitoshi Yamamoto$^b$}
\address{$^a$Department of Nuclear Physics, University of Madras,\\
Guindy Campus, Chennai 600 025, Tamil Nadu, INDIA\\
$^b$ Department of Physics and Astronomy, University of Hawaii,\\
2505 Correa Road, Honolulu, HI 96822, USA}

\date{\today}

\maketitle

\begin{abstract}
We study the inclusive charmless semileptonic decay of $\Lambda_b$,
$\Lambda_b \rightarrow X_u l \nu_l$, in the
framework of heavy quark expansion and find that the
rate is substantially enhanced by spectator effect.
We obtain $Br(b \rightarrow X_u l \nu_l)$ to be $1.13 \times 10^{-3}$ for
$|V_{ub}| = 3.3 \times 10^{-3}$. We discuss our result in light of the
recent LEP measurements of the charmless
semileptonic branching ratio where
such enhancement was not taken into account in
extracting $|V_{ub}|$.

\end{abstract}
%\twocolumn
\vskip0.5cm
Precise determination of the CKM matrix elements is fundamentally
an important task for the Standard Model. There are as many as
five of the nine elements of the CKM matrix that can be extracted
from the knowledge of the weak decays of beauty hadrons.
Particularly, $|V_{cb}|$ and $|V_{ub}|$ can be extracted in the framework
of heavy quark expansion (HQE) in a direct and fairly model independent way
\cite{neubert}.
With the measurements of the inclusive charmless semileptonic branching ratio of $b$
hadrons by the ALEPH \cite{aleph}, L3 \cite{L3} and DELPHI \cite{d}
collaborations at LEP, the
determination of $|V_{ub}|$ has attracted renewed interest recently.
The measured values of inclusive charmless semileptonic branching ratio of 
$b$ hadrons and the corresponding
values of $|V_{ub}|$ extracted are
\bea
  &&\begin{array}{rcl}
    Br(b \rightarrow X_u l\nu_l) &=& (1.73 \pm 0.55 \pm 0.55) \times 10^{-3},\\
    |V_{ub}| &=& (4.16 \pm 1.02)\times 10^{-3}.
  \end{array}  \hspace{0.7in} (ALEPH) \\
  &&\begin{array}{rcl}
    Br(b \rightarrow X_u l\nu_l) &=& (3.3 \pm 1.0 \pm 1.7) \times 10^{-3},\\
   |V_{ub}| &=& (6.0^{+0.8+1.4}_{-1.0-1.9} \pm 0.2) \times 10^{-3}.
  \end{array}  \hspace{0.9in} (L3) \\
  &&\begin{array}{rcl}
    Br(b \rightarrow X_u l\nu_l) &=& (1.57 \pm 0.35 \pm 0.48 \pm 0.27)\times10^{-3},\\
   |{V_{ub}/ {V_{cb}}}| &=& (0.103^{+0.011}_{-0.012} \pm 0.016 \pm 0.010).
  \end{array}  \hspace{0.24in} (DELPHI)
\eea
In these analyses, the measured quantity is
\be
{Br(b \rightarrow u e \nu) \over {Br(b \rightarrow c e \nu)}}
\propto \left| {V_{ub} \over {V_{cb}}} \right|^2\,.
\ee
This has an advantage of being free of many hadronic uncertainities
that occur in the non-leptonic decays. Also,
the determination of the CKM matrix elements from inclusive decays can in general be made
with less theoretical uncertainties than ones extracted from exclusive modes.
For semileptonic decays of baryons, however, there could still
be large spectator effects due to Pauli interference as pointed out
by Voloshin for $\Xi_c$~\cite{voloshin}.
To the inclusive charmless semileptonic branching ratio of
$b$ hadrons, the contribution of $\Lambda_b$ is about 10\%
with the rest coming from the $B$ mesons. There are many theoretical works
\cite{jin} which study the $B \rightarrow X_u l \nu_l$ to determine
$|V_{ub}|$. But not much has
been done on the inclusive charmless semileptonic decay of
$\Lambda_b$. In the analyses of ALEPH and L3, 
$\Lambda_b \rightarrow X_u l \nu_l$ is included in their sample,
but in that of DELPHI, the rejection of kaons and protons 
that was used in the selection criteria
reduces the contributions from $\Lambda_b$.

In this letter, we study the inclusive charmless semileptonic decay of
$\Lambda_b$. We find that the baryonic decay rate is larger
by a factor of about 1.36, due to spectator effects, than that of
the $b$ quark decay rate. We then discuss the correction factors
needed on $Br(b \rightarrow u l \nu_l)$ to account for the
spectator effect in $\Lambda_b \rightarrow X_u l \nu_l$.

The total rate of inclusively decaying beauty hadrons into a
charmless final state is given by the HQE\cite{neubert} as
\be
\Gamma(H_b) = {G_f^2 |V_{ub}|^2 m^5 \over {192 \pi^3}}
\left( 1 - {\lambda_1 - 3 \lambda_2 \over {2m^2}} - 2 {\lambda_2
\over {m^2}} + O\left(1 \over {m^3}\right)\right)
 \label{rate}
\ee
where with $m$ being the mass of the heavy quark, the term of
$O(1/m^0)$ corresponds to the free heavy quark
decay rate, the terms at $O(1/m^2)$ describe the motion of the
heavy quark inside the hadron ($\lambda_1$ = $-$0.5  $GeV^2$ for $B$ and
$-$0.43  $GeV^2$ for $\Lambda_b$) and the chromomagnetic
interaction due to the heavy quark spin projection
($\lambda_2$ = 0.12  $GeV^2$)
which vanishes for baryons except $\Omega_Q$ and
the third order term in $1/m$ is given by
\begin{center}
$C(\mu)\left<H|(\bar b \Gamma q)(\bar q \Gamma b)|H\right>$.
\end{center}
The operators in $\left<...\right>$, denoted as $\left<O_6\right>$ below,
are evaluated in \cite{arun}
by one of us for $B$ and $\Lambda_b$. The Wilson coefficients
are describing the spectator quark processes like Pauli
interference, weak annihilation and $W$-scattering
which is in baryons only.

\begin{figure}[t]
 \begin{center}
 \epsfig{figure=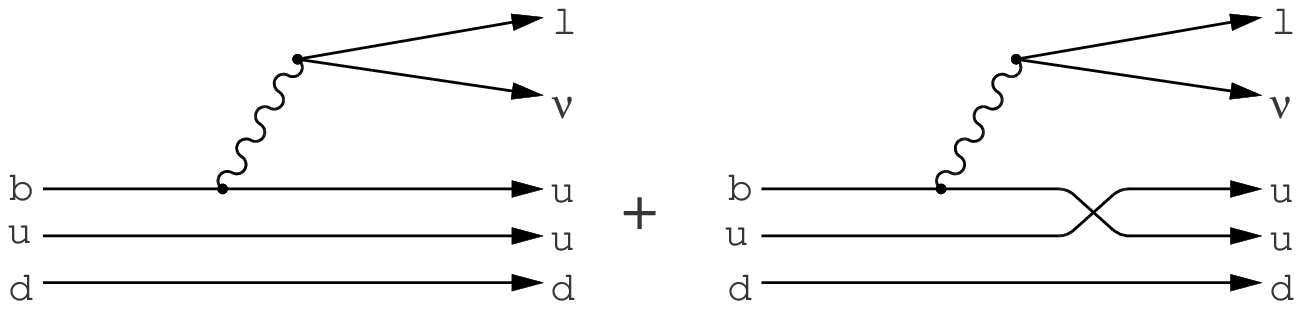, width=5.8in}
 \caption{The Pauli interference in the charmless
   semileptonic decay of $\Lambda_b$.} 
 \label{fg:diag}
 \end{center}
\end{figure}

In the mode $\Lambda_b \rightarrow X_u l \nu_l$, the
spectator effect is the constructive interference of the $u$ quark
of the final state with the $u$ quark in the initial hadron 
(see Fig.~\ref{fg:diag}) which
enhances considerably the decay rate. Otherwise,
without spectator effects, the baryonic decay rate is almost the same
as that of the heavy quark decay rate.
The decay rate due to spectator processes is given by
\be
 {\Delta \Gamma (\Lambda_b) = 48 \pi^2 \Gamma_0 
 {\left<O_6\right>_{\Lambda_b} \over {m^3}}}
\ee
where $\Gamma_0 = G_f^2 |V_{ub}|^2 m^5/192 \pi^3$. Using the
expectation values of the four-quark operators obtained in Ref. \cite{arun}
for $\Lambda_b$, $\left<O_6 \right>_{\Lambda_b} = 6.75 \times 10^{-2} GeV^3$,
we obtain the ratio
\be
  {{\Gamma(\Lambda_b \rightarrow X_u l \nu_l)} \over
  {\Gamma(b \rightarrow X_u l \nu_l)}} = 1.36.
  \label{ratio}
\ee
whereas at $O(1/m^2)$, the ratio is
$\Gamma(\Lambda_b \rightarrow X_u l \nu_l)/
\Gamma(b \rightarrow X_u l \nu_l) = 1.01$. It should be pointed out that
this enhancement has a large theoretical uncertainty which is difficult to estimate.
On the other hand, no such enhancement exists for 
$\Lambda \rightarrow X_c e \nu_e$ and thus the ratio of eq. (5)
is increased by 36\% for $\Lambda_b$. 
In passing, we note that
the hadronic $b \rightarrow u$
decay of $\Lambda_b$ is not substantially enhanced due to the 
cancellations of constructive $uu$ and destructive $dd$ interferences. 

The inclusive charmless semileptonic branching ratio of $b$ hadrons is
given by 
\be
Br(b \rightarrow X_u l \nu_l) = (1-f_{\Lambda_b}) 
\Gamma(B \rightarrow X_u l \nu_l) \tau(B) + f_{\Lambda_b}
\Gamma(\Lambda_b \rightarrow X_u l \nu_l)\tau(\Lambda_b)
\ee
In principle, $\Xi_b^{0,-1}$ can also be produced at LEP. However,
it is suppressed by $s\bar s$ production from vacuum relative to
$u\bar u$ or $d\bar d$ production. Furthermore, the neutral $\Xi_b$
which consists of $b$, $u$, and $s$ quarks has the same enhancement
factor as $\Lambda_b$ in the charmless semileptonic decay rate, even though
the charged $\Xi_b$ which is made of $bds$ quarks
does not have the corresponding enhancement.
We will thus assume in this analysis that all weakly decaying b-baryons
are $\Lambda_b$'s.
Thus using the above estimate with $f_{\Lambda_b}$ = 10\%,
$\tau(B)$ = 1.64 $ps$ and $\tau(\Lambda_b)$ = 1.24 $ps$, we obtain
the following value for inclusive charmless semileptonic branching ratio of $b$ hadrons:
\be
Br(b \rightarrow X_u l \nu_l)  =  1.16 \times 10^{-3}
\ee
If the spectator effects are
not included, then
\be
 Br(b \rightarrow X_u l \nu_l)  =  1.13 \times 10^{-3}
 \label{brbunsp}
\ee
In the above estimate in Eqs. (\ref{ratio}-\ref{brbunsp}), we have employed
$|V_{ub}| = 3.3 \times 10^{-3}$ and $m$ = 4.5 $GeV$.
Also, we have used the decay rate for
$B$ mesons as estimated at $O(1/m^2)$ using the Eq. (\ref{rate}),
$Br(B \rightarrow X_u l \nu_l) = 1.155 \times 10^{-3}$.
No spectator effects occur in $b \rightarrow u l \nu$ transition in $B$
mesons except the negligible weak annihilation in $B^-$.
Also, the experimental selection criteria used for  
the $X_u$ sytem mostly reject such final states.

The corrections on $|V_{ub}|$ due to the enhancement of 
$\Lambda_b \rightarrow X_u l \nu_l$ is about $-$1.3\%, and
with these corrections, the central values for $|V_{ub}|$ by
ALEPH and L3 becomes, as given by
$|V_{ub}| = \left({Br_{expt} / {Br_{theory}}}\right)^{1/2}0.0033$,
as
\bea
|V_{ub}|  & = & 4.03 \times 10^{-3},~~~(ALEPH)\\
 & = & 5.57 \times 10^{-3},~~~(L3)
\eea

In conclusion, we studied the inclusive charmless semileptonic
decay of $\Lambda_b$ and calculated the inclusive charmless semileptonic
branching ratio of $b$ hadrons.
Finally, we observe that
(1) the contribution of $\Lambda_b$ definitely influences
the branching ratio due to a constructive
Pauli intereference and (2) such spectator effects in $\Lambda_b$
decay has to be taken into account for a precise
determination of $|V_{ub}|$ which relies on chamless semileptonic
decays.

\acknowledgements
We thank the organisers of the 6th Workshop on High Energy Physics
Phenomenology (held at Chennai, India, during Jan. 3 - 15, 2000) for
their invitation where this work began. One of us (H. Y.) is supported by
the U.S. Department of Energy Grant \# DE-FG02-91ER40654,
whereas the other (S. A.) thanks Prof. P. R. Subramanian for useful discussions
and encouragement and Prof. Yasuhiro Okada for hospitality at KEK, Japan during
the preparation of this paper.

\end{document}